\documentclass[doublecol,linenumbers]{epl2}
\usepackage{graphicx}
\usepackage[english]{babel}
\usepackage{latexsym,amsmath,amscd,amssymb,graphicx,amsbsy,color,xcolor,amsfonts,amsthm}
\title{New metrics of a spherically symmetric gravitational field passing classical tests of General Relativity}
\subtitle{ Europhysics Letters \textbf{126} (2019) 29001 (corrected)}
\author{Yaakov Friedman and Shmuel Stav}
\institute{Jerusalem College of Technology\\Department of  Physics \\
P.O.B. 16031 Jerusalem 91160, Israel}


\abstract{
A general form of a  metric preserving all symmetries of a spherically symmetric gravitational field and angular momentum in spherical coordinates is obtained. Such metric may have $g_{01}(r)\neq 0$. The Newtonian limit uniquely defines $g_{00}(r)$. Geodesic motion under such metric exactly reproduces the precession of a planetary  orbit, periastron advance  of a binary, deflection of light and {Shapiro time delay} if the determinant of the time-radial parts of the metric is $-1$.  In this model, the total time for a radial round trip of light is as in the Schwarzschild model,  but it allows for light rays to have different  speeds propagating  toward or from the massive object. The value of $g_{01}(r) $ could be obtained by measuring these speeds.  All of these metrics do satisfy Einstein's field equations.}

\begin{document}
\maketitle

\section{Introduction}\label{intro}
$\;$
In \cite{Ein15}   A. Einstein proposed  to represent  planetary motion as geodesic motion with respect to a metric $g_{\mu\nu}$ on spacetime, which is spherically symmetric, asymptotically flat and also satisfies the ``equation of the determinant" $|g_{\mu\nu}|=1.$  He also assumed that $g_{0j}=g_{j0}=0,$ for $j=1,2,3$.
 The  Schwarzschild metric \cite{Sch} is of this type.

 Using spherical coordinates, we describe all possible metrics $g_{\mu\nu}$ on flat spacetime of a gravitational field of a non-rotating spherically symmetric body. Such metrics, which preserve all the symmetries of the problem and preserve angular momentum for geodesic motion (defined by Euler-Lagrange equations),  are characterized by $g_{00}(r), g_{01}(r)$ and $g_{11}(r)$. The classical limit  determines the $g_{00}(r)$ component of the metric. The allowed transformations preserving the metric are only spatial rotations, as in  \cite{Ein15}. This limits the allowed transformations under this model in comparison  to the GR model, and  we cannot apply arguments, like in  \cite{LL}, to show that the metric of such a field could  be transformed to diagonal form. Also, we do not want to assume a priori that the speed of light toward and from the massive object is the same, a property which was used in \cite{Rindler} to show that the off-diagonal components of the metric vanish. Thus, we do not assume  $g_{01}=0,$ which  is not implied by the symmetry of the problem.

 We will show that this model predicts the observed anomalous precession of Mercury's orbit, the  periastron advance of a binary,  gravitational lensing and {Shapiro time delay} if and only if  the determinant of the metric in two coordinates $(ct,r)$ is $-1$, as is assumed in \cite{Ein15}. This is true without specifying the components $g_{01}$  and $g_{11}$.

 The total time for a radial round trip of light is as in the Schwarzschild model,  so round trip experiments cannot distinguish between our model and $GR$.  But our model allows for light rays to have different  speeds propagating  toward or from the massive object. One of the metrics introduced here is analytic at all points except the origin. In this metric, the speed of light toward the object is always $c$, but the speed of light from the object decreases with the decrease of $r$ and becomes zero at the Schwarzschild radius.  All of these metrics {\bf do} satisfy Einstein's field equations.

 The results presented here are based on the ideas of  Relativistic Newtonian Dynamics \cite{FSS} and \cite{GN} applied to a general, spherically symmetric gravitational field.

\section{The spherical symmetric metric}

Consider a gravitational field generated by a spherically symmetric, non-rotating  mass $M.$ We define a metric on spacetime under which the motion of an object is a geodesic with respect to this metric.

 To define the spacetime, we place an imaginary observer far away from the sources of the field. Since the observer is not affected by the forces, we can assume that he measures space increments and time intervals as in  Minkowski space.  For convenience, we place the origin of our frame $K$ at the center of the symmetry of the field and use standard spherical coordinates $ct,r,\theta,\varphi$.

 { As known, (see for example \cite{HEL}, p. 197) a spherically symmetric stationary metric} in $K$ is of the form
 \begin{equation*}
   ds^2=g_{00}(r)c^2dt^2-g_{11}(r)dr^2- 2cg_{01}(r)dt dr
 \end{equation*}
\begin{equation}\label{genmet}
-l(r)r^2(d\theta^2+\sin^2\theta d\varphi^2).
\end{equation}
Since the force is static, the metric coefficients do not depend on $t$.  By spherical symmetry, the functions $g_{00},g_{01},g_{11}$ and $l$ cannot depend on
$\varphi$ or $\theta$. Since spatial rotation in spherical coordinates changes only $\varphi$ and $\theta$ and preserves the angular part of the metric, the metric (\ref{genmet}) is spherically symmetric.

 For points far removed from the sources, we assume, as usual,  that the metric is  the Minkowski metric. Hence,
 \begin{equation*}
  \lim_{r\rightarrow\infty}g_{00}(r)=\lim_{r\rightarrow\infty}g_{11}(r)=\lim_{r\rightarrow\infty}l(r)=1,
 \end{equation*}
\begin{equation}\label{normal}
\lim_{r\rightarrow\infty}g_{01}(r)=0.
\end{equation}

The trajectory of an object with mass $m$   is parameterized by proper time $d\tau=c^{-1}ds$. Its geodesic motion is obtained by optimizing with respect to the Lagrangian function $L(x,\dot{x})=mc\frac{ds}{d\tau}$ (see \cite{FSS}). {As it is shown in \cite{GN} p.3,  since the Lagrangian does not depend on $\varphi$ and  angular momentum is conserved  on any geodesic trajectory}, one obtains that $l(r)\equiv 1$. Thus, the metric is characterized by $g_{00}(r), g_{01}(r)$ and $g_{11}(r)$.

In what follows, we need the notion of a determinant of the metric, defined as follows. Restrict spacetime temporarily to the first two coordinates $(ct,r)$. In these coordinates, the matrix of the metric $g_{\alpha\beta}$ is
\begin{equation}\label{matr}
g_{\alpha\beta}=  \left(
    \begin{array}{cc}
      g_{00}(r) &- g_{01}(r) \\
     - g_{01}(r) &- g_{11}(r)\\
    \end{array}
  \right)
\end{equation}
Denote  by $-g$ the determinant of this matrix, then
\begin{equation}\label{det}
 g= g_{00}(r)g_{11}(r)+g_{01}(r)^2.
\end{equation}
The matrix of the inverse metric is
\begin{equation}\label{matrInv}
g^{\alpha\beta}= \frac{1}{g} \left(
    \begin{array}{cc}
      g_{11}(r) & -g_{01}(r) \\
       -g_{01}(r) &- g_{00}(r)\\
    \end{array}
  \right).
\end{equation}

\section{ Implication of the classical limit on the metric}

Consider radial motion. The trajectory of this  motion  is optimized with respect to the function $L(x,\dot{x})$, which  in this case is
\begin{equation}\label{Lr}
L\left(t,r,\dot{t},\dot{r}\right)=mc\sqrt{g_{00}(r)c^2\dot{t}^2-g_{11}(r)\dot{r}^2-2cg_{01}(r)\dot{t}\dot{r}},
\end{equation}
where the $\cdot$ denotes  differentiation by $\tau$. The Euler-Lagrange equation for the $r$ coordinate is
\begin{equation}\label{ELr}
  \frac{\partial L}{\partial r}-\frac{d}{d\tau}\frac{\partial L}{\partial \dot{r}}=0.
 \end{equation}

The $r$-momentum is $p_r=\frac{\partial L}{\partial \dot{r}}=-m(g_{11}(r)\dot{r} +cg_{01}(r)\dot{t})$,  its $\tau$ derivative is
\begin{equation}\label{rmomder}
 \frac{d}{d\tau}\frac{\partial L}{\partial\dot{r}}=-m\left( g'_{11}(r)\dot{r}^2 +g_{11}(r)\ddot{r} +cg'_{01}(r)\dot{t}\dot{r}+c g_{01}(r)\ddot{t}\right),
\end{equation}
and
\begin{equation}\label{dldr}
  \frac{\partial L}{\partial r}= \frac{m}{2}(g'_{00}(r)c^2\dot{t}^2-g'_{11}(r)\dot{r}^2-2cg'_{01}(r)\dot{t}\dot{r}).
\end{equation}
 Equation  (\ref{ELr}),  after cancelation of the  term $cg'_{01}(r)\dot{t}\dot{r}$ and dividing by $m/2$, becomes
\begin{equation}\label{ELRad}
g'_{00}(r)c^2\dot{t}^2+g'_{11}(r)\dot{r}^2+2g_{11}(r)\ddot{r} +2cg_{01}(r)\ddot{t}=0.
\end{equation}

We define now the function $g_{00}(r)$  from the Newtonian classical limit. Let $r_0$ be an arbitrary value of $r$. Consider the radial motion of an object whose velocity at $r_0$ is $\frac{dr}{dt}(r_0)=0$. Since $\dot{r}= \dot{t}\frac{dr}{dt},$ also  $\dot{r}(r_0)=0$.
From (\ref{genmet}) and  the definition of $d\tau$, we  have
\begin{equation}\label{tdot}
  \dot{t}=\frac{1}{\sqrt{g_{00}(r)-\frac{g_{11}(r)}{c^2} \left( \frac{dr}{dt}\right)^2-\frac{2g_{01}(r)}{c}\left( \frac{dr}{dt}\right)}},
\end{equation}
implying that
\begin{equation}\label{accr0}
\dot{t}(r_0)= \frac{1}{\sqrt{g_{00}(r_0)}}  \;\; \mbox{ and }\;\;       \ddot{r}(r_0)=\frac{1}{g_{00}(r_0)}\frac{d^2r}{dt^2}(r_0).
\end{equation}
Differentiating (\ref{tdot}) and substituting $r=r_0$ yields
\begin{equation*}
\ddot{t}(r_0)=\frac{1}{2}\frac{1}{g_{00}(r_0)\sqrt{g_{00}(r_0)}}\frac{2g_{01}(r_0)}{c\sqrt{g_{00}(r_0)}}\frac{d^2r}{dt^2}(r_0)
\end{equation*}
\begin{equation}\label{tddot}
  =\frac{g_{01}(r_0)}{cg_{00}(r_0)^2}\frac{d^2r}{dt^2}(r_0).
\end{equation}
Substituting this into (\ref{ELRad}) and multiplying this equation by $g_{00}(r_0)^2,$ we obtain
\begin{equation*}c^2g_{00}(r_0)g'_{00}(r_0)+2\left( g_{00}(r_0)g_{11}(r_0)+g_{01}^2(r_0) \right)\frac{d^2r}{dt^2}(r_0)=0,\end{equation*}
and, using (\ref{det}), we have
\begin{equation}\label{accelr0}
  \frac{d^2r}{dt^2}(r_0)=-\frac{c^2}{2g}g_{00}(r_0)g'_{00}(r_0).
\end{equation}

Let $U(r)=-GM/r$ denote the classical Newtonian gravitational potential of this field. The Newtonian radial acceleration in tensorial form is $\frac{d^2r}{dt^2}=m^{-1}g^{1j}U,_j$ (see \cite{Itin}). Using (\ref{matrInv}) and that the gradient of $U(r)$ is in the radial direction, we have
\begin{equation}\label{2Newt}
   \frac{d^2r}{dt^2}(r_0)=-\frac{1}{m}\frac{g_{00}(r_0)}{g}U'(r_0).
\end{equation}
Comparing this to (\ref{accelr0}) and using that $r_0$ was arbitrary, we obtain
\begin{equation}\label{f-eq}
 g'_{00}(r)=\frac{2}{mc^2}U'(r).
\end{equation}
Integrating and using (\ref{normal}), this implies that
\begin{equation}\label{ffinal}
  g_{00}(r)=1-u(r),\;\;\; u(r)=-\frac{2U(r)}{mc^2}=\frac{r_s}{r},
\end{equation}
where $r_s=\frac{2GM}{c^2}$ is the  Schwarzschild radius.

\section{ Precession of planetary orbits}

Now we return to general motion (not radial) of an object in the spherically symmetric gravitational field. The motion is by a geodesic with respect to the metric
\begin{equation*}
 ds^2=g_{00}(r)c^2dt^2-g_{11}(r)dr^2-2 cg_{01}(r)dt dr
\end{equation*}
\begin{equation}\label{metric}
-r^2(d\theta^2+\sin^2\theta d\varphi^2),
\end{equation}
with $g_{00}(r)$ defined by (\ref{ffinal}).
From the symmetry of the problem, it follows that the trajectory is in a plane passing through the center of the gravitational field. This plane is determined by the initial position of the object and its initial velocity. Thus, without loss of generality we
will assume that the motion is in the plane $\theta=\pi/2$.

To be able to handle motion of both massive objects and massless particles, introduce a symbol $\varepsilon$ with value 1 for massive objects and 0 for massless particles.
 Since, for massless particles, the line interval $ds$ defined by (\ref{metric}) is zero, dividing (\ref{metric}) by $d\tau^2$ we obtain
\begin{equation}\label{norm4vel}
  c^2\varepsilon=g_{00}(r)c^2\dot{t}^2-g_{11}(r)\dot{r}^2- 2cg_{01}(r)\dot{t}\dot{r}-r^2\dot{\varphi}^2.
\end{equation}
  Since our metric (\ref{metric}) is independent of $\varphi, $ the momentum corresponding to this variable is conserved, implying
\begin{equation}\label{angMom}
  r^2\dot{\varphi}=J,
\end{equation}
where $J$ has the meaning of angular momentum per unit mass.

 Since our metric (\ref{metric}) is also independent of  $t,$ the momentum
 \begin{equation}\label{pt}
  p_t=g_{00}(r)c\dot{t}-g_{01}(r)\dot{r}
 \end{equation}
is conserved.  Using (\ref{norm4vel}), (\ref{angMom}) and (\ref{det}) we obtain
\begin{equation*}p_t^2=g_{00}^2(r)c^2\dot{t}^2-2cg_{00}(r)g_{01}(r)\dot{t}\dot{r}+g^2_{01}(r)\dot{r}^2=\end{equation*}
\begin{equation*}
 g_{00}(r)( c^2\varepsilon+g_{11}(r)\dot{r}^2+ 2cg_{01}(r)\dot{t}\dot{r}+r^2\dot{\varphi}^2)
\end{equation*}
\begin{equation*}-2cg_{00}(r)g_{01}(r)\dot{t}\dot{r}+g^2_{01}(r)\dot{r}^2=\end{equation*}
\begin{equation*} c^2 \varepsilon g_{00}(r)+(g_{00}(r)g_{11}(r)+g^2_{01}(r)) {\dot{r}^2}+g_{00}(r)\frac{J^2}{r^2}\end{equation*}
\begin{equation*}
  =  c^2\varepsilon g_{00}(r)+g\dot{r}^2+g_{00}(r)\frac{J^2} {r^2}.
\end{equation*}
Using (\ref{ffinal}),  this implies that
\begin{equation}\label{ConsEq}
g\dot{r}^2=-(1-u)\left(c^2\varepsilon+\frac{J^2} {r^2}\right)+p_t^2.
\end{equation}

We will solve the last equation for $u(\varphi)$ on the trajectory. From (\ref{ffinal}) and (\ref{angMom}), it follows that
\begin{equation*}u'= \frac{du}{d\varphi}=-\frac{r_s}{r^2}\frac{dr}{d\varphi}=-\frac{r_s}{r^2}\frac{\dot{r}}{\dot{\varphi}}=-\frac{r_s}{J}\dot{r}\end{equation*}
and $\dot{r}=-\frac{J}{r_s}u'$.  Substituting this into (\ref{ConsEq}), using that $\varepsilon=1,$  multiplying by $\frac{r_s^2}{J^2}$ and denoting $2\mu=\frac{c^2r_s^2}{J^2}$, we obtain
\begin{equation}\label{precesseq}
  g{u'}^2=(u-1)(u^2+2\mu)+p_t^2r_s^2/4J^2=u^3-u^2 +2\mu u +const.
\end{equation}
Consider now the case when the orbit is bounded. In this case, there are two points on the orbit corresponding to the perihelion and aphelion on the trajectory, where $u'$ vanishes. These are two of the roots of the cubic polynomial in
$u$ on the right side of the above equation. From this, by standard arguments, one shows that the solution is a precessing ellipse.  The precession is the one predicted by $GR$ if and only if $g=1$.

As shown in \cite{FSBin},  the same derivation leads to the  correct formula for the periastron advance of a binary if $g=1$.

\section{Gravitational lensing and the Shapiro time delay}

Gravitational lensing and the Shapiro time delay (or gravitational time delay) describe the deflection of a light ray and the slowing of a light pulse ($\varepsilon=0$) as it moves from a point $A$ to a point $B$  in the gravitational potential of a spherically symmetric massive object of mass $M$. For light propagation,  equation (\ref{ConsEq}) becomes
\begin{equation}\label{shap1}
g\dot{r}^2=-(1-u)\frac{J^2} {r^2}+p_t^2.
\end{equation}

Consider now the trajectory $r(\varphi)$ of the light ray. Using (\ref{angMom}), we obtain $\dot{r}=\frac{dr}{d\varphi}\frac{J}{r^2}$.  Substituting this into the above equation and dividing by $p_t^2$ yields
\begin{equation}\label{shap2}
g\left(\frac{dr}{d\varphi}\frac{J}{p_t r^2}\right)^2=-(1-u)\frac{J^2} {p_t^2r^2}+1.
\end{equation}
Denote by $r_0$ the position on the trajectory closest to the center of the massive object. Then $\frac{dr}{d\varphi}(r_0)=0,$ and, from the above,
\begin{equation}\label{bdef}
  \frac{J}{p_t}=\frac{r_0}{\sqrt{1-u(r_0)}}=b.
\end{equation}

To obtain the formula for gravitational lensing, substitute this  into (\ref{shap2}), which yields
\begin{equation}g\left(\frac{r_0}{r^2}\frac{dr}{d\varphi}\right)^2+\left(1-\frac{r_s}{r}\right)\frac{r_0^2}{r^2}=1-\frac{r_s}{r_0}.\end{equation}
For any angle $\varphi$ on the trajectory, one may associate an angle $\alpha(\varphi)$ for which $r(\varphi)=\bar{r}(\alpha)$, where $\bar{r}(\alpha)=\frac{r_0}{\sin \alpha}$ is the
straight-line approximation of the trajectory at the point $P$, chosen to be the $x$ direction.
This suggests the substitution $r=\frac{r_0}{\sin\alpha}$,  which implies $\frac{dr}{d\varphi}=-\cos\alpha \frac{r^2}{r_0}\frac{d\alpha}{d\varphi}$ and
\begin{equation}\label{dpda}
\frac{d\varphi}{d\alpha}=\sqrt{g} \left( 1-\frac{r_s}{r_0}\left(\sin \alpha+\frac{1}{1+\sin\alpha}\right)\right)^{-1/2}.
\end{equation}
If $g=1,$ this equation is the same as obtained  for gravitational lensing in \cite{FSLens}.

Thus, if  points $A$ and $B$ are very remote from the massive body ($\alpha_A\approx \pi, \alpha_B\approx 0$) and  $r_s/r_0\ll 1$, the weak deflection angle becomes
\begin{equation}\label{Lens}
 \delta\phi\approx\frac{2r_s}{r_0}=\frac{4GM}{c^2 r_0},
\end{equation}
which is identical to the angle given by Einstein's formula for weak gravitational lensing using $GR$ (\cite{MTW,Rindler}).

To obtain the formula for the Shapiro time delay, using (\ref{angMom}),  (\ref{pt}) and (\ref{bdef}), we obtain
\begin{equation*} \frac{d\varphi}{dt}=\frac{\dot{\varphi}}{\dot{t}}=\frac{Jg_{00}(r)c}{r^2(p_t+g_{01}(r)\dot{r})}=\frac{bg_{00}(r)c}{r^2(1+g_{01}(r)\dot{r}/p_t)}.\end{equation*}
Formula (\ref{shap1})  yields
\begin{equation*}\frac{\dot{r}}{p_t}={\pm}\frac{1}{\sqrt{g}}\sqrt{1-g_{00}(r)\frac{b^2}{r^2}},\end{equation*}
{where the sign is chosen depending of $\dot{r}$.}
Thus,
\begin{equation*} \frac{d\varphi}{dt}=\frac{bg_{00}(r)c}{r^2\left(1{\pm}\frac{g_{01}(r)}{\sqrt{g}}\sqrt{1-g_{00}(r)\frac{b^2}{r^2}}\right)}\end{equation*}
and
\begin{equation*}cdt=\frac{r^2}{bg_{00}(r)}\left(1{\pm}\frac{g_{01}(r)}{\sqrt{g}}\sqrt{1-g_{00}(r)\frac{b^2}{r^2}}\right)d\varphi.\end{equation*}

{ For a signal traveling from $A$ to $B$ and back, we integrate each point twice, one time when $\dot{r}>0$ and the second time when  $\dot{r}<0$. Thus, in the above formula, the term based on $g_{01}(r)$ is once added and once subtracted. This imply that the delay} is exactly as for the Schwarzschild metric.
Thus, the Shapiro time delay for a signal traveling from $A$ to $B$ and back is approximately
\begin{equation}\label{TimeDelf}
  r_s\ln \frac{4x_B|x_A|}{r_0^2},
\end{equation}
which is the known formula for the Shapiro time delay (\cite{MTW,Rindler}), confirmed by several experiments.

\textbf{Conclusion.} If the determinant of the metric $g$ defined by (\ref{det}) is equal 1 and $g_{01}(r)\ll 1$,  the geodesic motion with respect to  the metric (\ref{metric}), satisfying the Newtonian limit expressed by (\ref{ffinal}),  passes all classical tests of $GR$.

\section{ Velocity of the light in the radial direction}

For light propagating in the radial direction of a spherically symmetric gravitation field, we will denote its speed  at $r$ in the inertial frame $K$ by $v_\uparrow (r),$ if the light moves away from the source of the field, and by $v_\downarrow (r)$ if the light moves toward the source.  We are not assuming that these speeds are the same. Since for light $ds=0$, these velocities satisfy
\begin{equation*}g_{00}(r)c^2-g_{11}(r)v^2-2cg_{01}(r)v=0,\end{equation*}
and using (\ref{det}), this implies  that
\begin{equation}\label{2 velocities}
 v_\uparrow (r)=\frac{c}{g_{11}(r)}(\sqrt{g}-g_{01}(r)) ,\; v_\downarrow (r)=\frac{c}{g_{11}(r)}(\sqrt{g}+g_{01}(r)).
\end{equation}
Thus, by measuring the values of $ v_\uparrow (r)$ and  $v_\downarrow (r)$  and assuming $g=1$, one is able to identify the full metric.

Note that from (\ref{normal}),  it follows that as  $r$ approaches  infinity, both speeds become the speed of light $c$ in an inertial frame, as expected.  Moreover, at radius $r_{\tilde{s}}$, when
\begin{equation}\label{radS}
  g_{01}(r_{\tilde{s}})=\sqrt{g},
\end{equation}
 there is no light propagating  away from the center of the field at any point with $r\leq r_{\tilde{s}}$.

Let us check the time $T$ that it takes for light to go radially from $r_1$ to $r_2>r_1$ and return:
\begin{equation*}
  cT=\int_{r_1}^{r_2} \frac{dr}{v_\uparrow}+\int_{r_1}^{r_2} \frac{dr}{v_\downarrow}
\end{equation*}
\begin{equation*}=\int_{r_1}^{r_2}\left( \frac{1}{\sqrt{g}-g_{01}(r)}+ \frac{1}{\sqrt{g}+g_{01}(r)} \right) g_{11}(r)dr\end{equation*}
\begin{equation*}= \int_{r_1}^{r_2} \frac{2g_{11}(r)\sqrt{g}}{g-g^2_{01}(r)}dr=2\int_{r_1}^{r_2} \frac{\sqrt{g}}{g_{00}(r)}dr. \end{equation*}
Since $g_{00}(r)$ defined by (\ref{ffinal}) is the same as in the Schwarzschild metric, the time of a round trip for light  is the same as in  the Schwarzschild metric if the determinant $g=1$.

\section{ Example of an analytic metric of a  spherically symmetric gravitational field}

Define a metric of a gravitational field of a spherically symmetric static object of mass $M$ positioned at the origin as
\begin{equation*}
  ds^2=(1-u(r))c^2dt^2-(1+u(r))dr^2- 2cu(r)dt dr
\end{equation*}
 \begin{equation}\label{anmet}
-r^2(d\theta^2+\sin^2\theta d\varphi^2),
\end{equation}
where $u(r)= -\frac{2GM}{rc^2}=\frac{r_s}{r}.$
In the first two coordinates $(ct,r)$ the matrix of the metric $g_{\alpha\beta}$ is
\begin{equation}\label{an matr}
g_{\alpha\beta}=  \left(
    \begin{array}{cc}
      1-u(r) & -u(r) \\
       -u(r) &-(1+u(r))\\
    \end{array}
  \right).
\end{equation}

This metric is analytic everywhere except the origin. Since for this metric $g=1$ and $g_{01}(r)\ll 1$, the  geodesic motion with respect to this metric passes all classical tests of $GR$. Note that the radius $r_{\tilde{s}}$, defined by (\ref{radS}) is the known Schwarzschild radius $r_s$. The two directional radial light velocities are
 \begin{equation}\label{an2 velocities}
 v_\uparrow (r)=\frac{1-u(r)}{1+u(r)}c  \;\;\mbox{  and  }\;\; v_\downarrow (r)=c,
\end{equation}
showing that for this model the speed of light $ v_\uparrow$ vanishes at the Schwarzschild radius, while $v_\downarrow $ is not affected by the gravitational field. This metric coincides with the Schwarzschild metric in Eddington–Finkelstein coordinates \cite{Edd}, \cite{Fin}. However, the interpretation of the coordinates in our model and in the Eddington–Finkelstein are different

\section{  Einstein's field equation for non-diagonal solution}

Assume that our metric is not diagonal and the determinant $g=1$. We can write the metric in spherical coordinates as
\begin{equation}\label{MetSpher}
  g_{\alpha\beta}=\left(
                    \begin{array}{cccc}
                      1-u & g_{01}(r) & 0 &0 \\
                      g_{01}(r) & \frac{g^2_{01}(r)-1}{1-u} & 0 & 0 \\
                      0 & 0 &- r^2 & 0 \\
                      0 &0 & 0&- r^2\sin^2\theta \\
                    \end{array}
                  \right),
\end{equation}
for $u(r)=r_s/r$ as in (\ref{ffinal}). The inverse metric is
\begin{equation}\label{InvMetSpher}
  g^{\alpha\beta}=\left(
                    \begin{array}{cccc}
                     \frac{1-g^2_{01}(r)}{1-u} & g_{01}(r) & 0 &0 \\
                      g_{01}(r) &u-1& 0 & 0 \\
                      0 & 0 & -r^{-2} & 0 \\
                      0 &0 & 0& -r^{-2}\sin^{-2}\theta \\
                    \end{array}
                  \right).
\end{equation}

Einstein's empty-space field equations outside the gravitating object, ignoring cosmological expansion, is
\begin{equation}\label{Einst}
  R_{\alpha\beta}=0,
\end{equation}
where
\begin{equation}\label{curv}
  R_{\alpha\beta}=\partial_\rho\Gamma_{\alpha\beta}^\rho-\partial_\beta\Gamma_{\rho\alpha}^\rho+\Gamma_{\rho\lambda}^\rho\Gamma_{\beta\alpha}^\lambda-\Gamma_{\beta\lambda}^\rho\Gamma_{\rho\alpha}^\lambda
\end{equation}
and
\begin{equation}\label{gamma}
  \Gamma_{\alpha\beta}^\rho =\frac{1}{2}g^{\rho\lambda}(g_{\lambda\alpha,\beta}+g_{\lambda\beta,\alpha}-g_{\alpha\beta,\lambda}).
\end{equation}
Using computer algebra, one show that (\ref{Einst}) is  satisfied for all $\alpha,\beta$.

\section{Discussion}

In his work \cite{Ein15}  on the motion of the perihelion of Mercury,  A. Einstein proposed (in current terminology) to represent  planetary motion as geodesic motion with respect to a metric $g_{\mu\nu}$ on spacetime, which is spherically symmetric, asymptotically flat and also satisfies  the ``equation of the determinant"
\begin{equation}\label{edet}
  |g_{\mu\nu}|=1.
\end{equation}
He also assumed that $g_{0j}=g_{j0}=0,$ for $j=1,2,3$. Einstein posed a problem to find a metric satisfying all these requirements. He showed that such a metric  leads to  Newton's second law in the first approximation, and the second approximation correctly reproduces the known anomaly in the motion of the perihelion of Mercury. The  metric  found  by K. Schwarzschild \cite{Sch} was such a metric.

 If we use spherical coordinates $(t,r,\phi,\theta)$ with respect to an inertial lab frame with the origin at the center of the spherically symmetric massive object, we may assume that $g_{01}\neq0$ and preserve the spherical symmetry of the field. We have shown that the classical limit defines $g_{00}$, as given by (\ref{ffinal}). If the ``equation of the determinant" (\ref{edet}) (or in our notation $g=1$) is  satisfied, the model reproduces the known anomalous precession of the perihelion of Mercury, periastron advance  of a binary, the deflection of light and {Shapiro time delay.  Moreover, the {orbits} of massive objects and massless particles are exactly the same in our model as in  $GR$. For massive objects this follow  from (\ref{precesseq})  and for massless particles from (\ref{shap1}).  Also the delay at round trip of light is the same as in $GR$.  Thus, any PPN parameters, based on measured trajectories and the Shapiro time delay, will be the same for our model and $GR$. }

{There is nowadays a great interest in alternative gravity theories, motivated by the possibility of explaining the flattering  of the rotation curves of galaxies without the need  of dark matter and energy and the accelerated expansion of the Universe. Unfortunately, for circular orbits our off-diagonal term does not produce corrections to $GR$ orbital velocities, since when $\dot{r}=0$ our formula for $\dot{t}$ is the same as in GR.  Thus, to handle the flattering we may need a modification of $g_{00}$ and $g_{11}$,  as it is done in \cite{BM} .}

 In this model, the total time for a radial round trip of light is as in the Schwarzschild model,  but unlike the Schwarzschild model,  it allows for light rays to have different  speeds propagating  toward or from the massive object. Measuring these speeds allows to identify the components of the metric.
We also presented an analytic non-diagonal metric with interesting properties of light propagation. This metric coincides with the Schwarzschild metric in Eddington–Finkelstein coordinates. It is also the metric in Synge's interpretation \cite{synge} of the Whitehead theory of gravitation \cite{White}.  All these  metrics satisfy Einstein's field equation.
\vskip0.3cm
 We  wish  to thank Dr. Tzvi Scarr for corrections and  { the referees for their constructive comments.}


\begin{thebibliography}{0}
\bibitem{Ein15} \Name{A. Einstein} \REVIEW{Sitzungsber. Preuss. Akad. Wiss., Phys. Math. Kl.} {778} {1915}{}
\bibitem{Sch} \Name{ K. Schwarzschild} \REVIEW{Sitzungsber. Preuss. Akad. Wiss., Phys. Math. Kl. } {189 } {1916} { }
\bibitem{LL} \Name{ L.  Landau \and E . Lifshitz } \Book{The Classical Theory of Fields.  Course of Theoretical Physics Vol. 2} \Publ{Pergamon, Addison-Wesley} \Year{1971}
\bibitem{Rindler} \Name{W. Rindler} \Book{Relativity, Special, General and Cosmological} \Publ{Oxford} \Year{2001}
\bibitem{FSS} \Name{Y. Friedman, T. Scarr \and J.M. Steiner} \REVIEW{Int. J. Geom. Meth. Mod. Phys.} {16}{2019}{1950015}
{\bibitem{GN}   \Name{Y. Friedman \and T. Scarr} \REVIEW{Europhys. Lett.} { 125} {2019} { 49001}}
{\bibitem{HEL} \Name{ M. P. Hobson, G. Efstathiou \and A. N. Lasenby} \Book{General Relativity} \Publ{Cambridge} \Year{2007}}
\bibitem{Itin}  \Name{Y. Itin} \REVIEW{International Journal of Geometric Methods in Modern Physics} {15} {2018} {1840002}
\bibitem{FSBin} \Name{Y. Friedman, S. Livshitz \and J.M. Steiner} J.M.: \REVIEW{Europhys. Lett.} {116} {2016} {59001-59006}
\bibitem{FSLens}  \Name{Y. Friedman \and J.M. Steiner} \REVIEW{Europhys. Lett.} {117} {2017} {59001}
\bibitem{MTW} \Name{C. W. Misner, K. S. Thorne \and J. A. Wheeler} \Book{Gravitation}  \Publ{Freeman and co.} \Year{1973}
{\bibitem{BM} \Name{O. Bertolam \and A. Martins} \REVIEW{Phys. Rev. D} {85} {2012} {024012}}
\bibitem{Edd}  \Name{A.S. Eddington} \REVIEW{Nature} {113} {1924}{2832}
\bibitem{Fin} \Name{ D. Finkelstein} \REVIEW{Phys. Rev.} {110}{1958}{965}
\bibitem{synge} \Name{ J. L. Synge} \REVIEW{Proc. R. Soc. London, Ser. A.} {211}{1952}{303}
\bibitem{White} \Name{A. N. Whitehead} \Book{The Principles of Relativity}  \Publ{Cambridge} \Year{1922}
\end{thebibliography}
\end{document}